\documentclass[12pt]{article}
\usepackage{amsfonts}
\def\del{\partial}
\def\Ds{D\!\!\!\! /}

\def\B{\mathcal{B}}
\def\X{\mathcal{X}}

\def\vD{\vec{D}}
\def\vt{\vec{\tau}}
\def\kok{\sqrt{2}}
\def\yarim{{{1}\over{2}}}

\def\a{\alpha}

\def\d{\delta}
\def\e{\epsilon}

\def\la{\lambda}

\def\x{\xi}

\def\adot{\dot{\alpha}}

\def\labar{\bar{\lambda}}

\def\xbar{\bar{\xi}}

\def\bs{\bar{s}}
\def\ts{\tilde{s}}

\def\be{\begin{equation}}
\def\ee{\end{equation}}
\def\bea{\begin{eqnarray}}
\def\eea{\end{eqnarray}}

\def\kok{\sqrt{2}}
\def\yarim{{{1}\over{2}}}

\begin{document}

%

\thispagestyle{empty}
\begin{center}
{\LARGE{ Construction of TYM via Field Redefinitions }}\footnote{Talk given at the International Workshop ”Supersymmetries and Quantum Symmetries” (SQS’05),Dubna, 27-31 July 2005.}\\
\vskip1cm
{\Large Kayhan \"{U}lker}\footnote{\small{e-mail: kulker@gursey.gov.tr}}
\vskip2em
{\sl Feza G\"{u}rsey Institute,}\\
{\sl \c{C}engelk\"{o}y, 81220, \.{I}stanbul, Turkey}\\
\vskip1.5em
\end{center}
\vskip1.5cm
%

\begin{abstract}
By constructing a nilpotent  extended BRST operator $\bs$ that involves the N=2 global supersymmetry
transformations of one chirality, we find exact field redefinitions that allows to construct the Topological Yang Mills Theory from the ordinary Euclidean N=2 Super Yang Mills theory in flat space. We also show that the given field redefinitions  yield the  Baulieu-Singer formulation of Topological Yang Mills theory when after an instanton inspired truncation of the theory is used. 
\end{abstract}

\section{Introduction}
Topological Yang-Mills (TYM) theory was first constructed by Witten \cite{witten} in 1988 as the twisted version of Euclidean N=2 Super Yang-Mills (SYM) theory in order to study the topological invariants of four-manifolds. Soon after \cite{witten}, it was shown by Baulieu and Singer that TYM can be fully obtained as a pure gauge fixing term  (i.e. as an exact BRST term) \cite{bs}. 

Moreover, N=2 SYM and TYM are also intertwined together when physical calculations are considered. For instance, the instanton calculations of N=2 SYM by using semi-classical approximation \cite{dkm} and the ones of TYM, where no approximation is needed to perform these calculations \cite{bftt2} give the same result. Since some position independent correlators exist in supersymmetric gauge theories \cite{am}, one can interpret this result that a subset of correlators of N=2 SYM coincide with a subset of the observables in TYM \cite{bftt2}. The non-renormalition theorems of N=2 SYM can also be proved by using twisted version \cite{sorlec}. As a consequence, one may conclude that twisting can be thought as a variable redefinition in flat space \cite{fb}.

Therefore, two questions are in order: First of all, is it also possible to write the action of N=2 SYM as an exact term like the twisted version of the theory and second, if twisting can be thought as really a variable redefinition in flat space, is it possible to find these field redefinitions explicitly?

The answers to both of the above questions are found to be affirmative \cite{ben} and the strategy to find these answers is to use the BRST formalism (also called BV or field-antifield formalism \cite{bv,gps}) that is extended to include global supersymmetry (SUSY). \cite{bon,w,mag,mpw,bra1}.

\section{Extended BRST  transformations and\\N=2 SYM action as an exact term}
The off-shell Euclidean N=2 SYM action \cite{gsw} is given as\footnote{Our conventions are explained in Ref.\cite{ben} in detail.},
\bea
I&=&Tr \int d^4 x (\frac{1}{4} F_{\mu\nu} F_{\mu\nu}
+\frac{1}{8}\e_{\mu\nu\la\rho}F_{\mu\nu}F_{\la\rho} - \la^i D \!\!\!\! / \labar_i  + M D_{\mu} D_{\mu}
N-\nonumber\\
&&\quad -\frac{i\kok}{2}(\la_i [\la^i , N ] +\labar^i [\labar_i , M ]) -\yarim [M,N ]^2 + \yarim\vD . \vD)
\eea
where the (anti-hermitian) gauge field $A_\mu  $ and the scalar fields $M\, ,\, N $ are singlets, the Weyl spinors $\la_{i\a} \,\,\labar^{i}_{\adot}$ are doublets and the auxiliary field $\vD $ is a triplet under the $SU(2)_R$ symmetry group.

Since, the action is translation, gauge and N=2 SUSY invariant one can define an extended BRST symmetry \cite{mag}:
\be
s=s_0 -i \x^i Q_i -i \xbar_i \bar{Q}^i -i \eta^{\mu}\del_{\mu} 
\ee
where $s_0$ is the ordinary BRST transformations, $Q_i \, ,\, \bar{Q}^i $ are chiral and antichiral parts of N=2 SUSY transformations and $\x^{i\a} \, ,\, \xbar_{i\adot} $ and $\eta _{\mu}$ are the constant \textit{commuting} chiral, antichiral SUSY ghosts and constant imaginary anticommuting translation ghost respectively. By choosing $s$-transformations of the ghosts suitably, the extended BRST operator $s$ becomes nilpotent \cite{mag} and one can construct a cohomology problem.

On the other hand, from the definition of $s$ it is still possible to derive another nilpotent operator  by using a suitable filtration of global ghosts \cite{psbook}. We choose this filtration to be
\be
\mathcal{N} = \xbar_{i\adot} \frac{\d}{\d \xbar_{i\adot}} + \eta_{\mu} \frac{\d}{\d \eta_{\mu}} \quad ;\quad s =
\sum{s^{(n)}} \quad ,\quad [\mathcal{N},
s^{(n)}]=ns^{(n)},
\ee
so that  the zeroth order in the above expansion is an operator that includes ordinary BRST and chiral SUSY on the space of the fields of the N=2 vector multiplet $M ,N ,A_\mu ,\la_i ,\labar^i,\vD$
\be
\bs:= s^{(0)} = s_0 -i \x^i Q_i\qquad,\qquad\bs ^2 =0.
\ee
The cohomology of $s$ is isomorphic to a subset of the cohomology of the filtered operator $\bs$ \cite{psbook}. The $\bs$ transformation of the fields are given as,
\bea
\bs A_{\mu}&=&  D_{\mu}c - \x_i e_{\mu}\labar^i \quad,\quad\bs M = - [c, M ] + i\kok \x^i\la_i  \quad,\quad\bs N = - [c, N ]\nonumber\\ 
\bs \la_i &=& - \{c,\la_i \} - e_{\mu\nu}\x_i F_{\mu\nu} +\x_i [M ,N ] + \vt_i ^j \x_j.\vD \quad,\quad\bs \labar^i = - \{ c,\labar^i \}  +i \kok \bar{e}_{\mu}\x^i D_{\mu} N \nonumber\\
\bs \vD &=& - [c, \vD] + \vt_i ^j(\x_j e_\mu D_\mu \labar^i  + i \kok\x^i [\la_j ,N])\nonumber \\
\bs c &=& - \frac{1}{2}\{ c,c\} + i\kok\x_i\x^i N \quad,\quad\bs \eta_{\mu}  = \bs \x_i = \bs \xbar_i = 0
\eea

Since, the actions of SYM theories can be represented as chiral (or antichiral) multiple supervariations of lower dimensional gauge invariant field polynomials \cite{ben1}, it is straightforward to assume that the action can also be written as an $\bs$ exact term of a gauge invariant field polynomial which is independent of Fadeev-Popov ghost fields\footnote{In other words, we assume that the action can be chosen to be a trivial element of equivariant cohomology of $\bs$.  See for instance Ref.s\cite{sorlec} and the references therein.},
\be
I=\bs \Psi.
\ee
It is clear that $\Psi$, the so called gauge fermion in BV formalism, has negative ghost number, $Gh(\Psi )=-1$. However, since no fields with negative ghost number has been introduced and since we have chosen the gauge fermion to be free of Fadeev-Popov ghosts, the only way to assign a negative ghost number to $\Psi$ is to choose $\Psi$ to depend on the negative powers of the global SUSY ghosts: 
\be
\Psi=\frac{1}{\x _k \x^k} \x^i \int d^4 x \psi_i
\ee
where $\psi_i ^\a $ is a dimension $7/2$  fermion that is made from the fields of the N=2 vector multiplet. The most general such gauge fermion that is covariant in its Lorentz, spinor and $SU(2)_R$ indices is easy to find:
\be
\Psi_E = \frac{1}{\x _k \x^k} Tr \int d^4 x  (\yarim \x ^i  \la_i  [M ,N ] -\yarim \x^i \vt _i ^j \la_j .
\vD - \frac{1}{2} \x^i e_{\mu\nu}\la_i  F_{\mu\nu} - \frac{i\kok}{2} M \x^i e_\mu D_\mu \labar_i ) .
\ee
The coefficients of the terms in $\Psi$ are fixed in order that the $\bs$ variation of $\Psi$ is free of chiral ghosts:
\bea
I_E&=&\bs\Psi_E\nonumber\\
&=&Tr \int d^4 x (\frac{1}{4} F_{\mu\nu} F_{\mu\nu}
+\frac{1}{8}\e_{\mu\nu\la\rho}F_{\mu\nu}F_{\la\rho} - \la^i D \!\!\!\! / \labar_i  + M D_{\mu} D_{\mu}
N\nonumber\\
&&\qquad\qquad\qquad
-\frac{i\kok}{2}(\la_i [\la^i , N ] +\labar^i [\labar_i , M ]) -\yarim [M,N ]^2 + \yarim\vD . \vD)
\eea
This is exactly the  N=2 supersymmetric Euclidean action, that is constructed by Zumino \cite{zumino}, up to the topological term $\e_{\mu\nu\la\rho}F_{\mu\nu}F_{\la\rho}$ and the auxiliary term $\yarim\vD . \vD$.

Here, we should remark that, the action belongs to the trivial cohomology of $\bs$ and therefore to that of the complete operator $s$, if and only if the functional space where $s$ is defined is the polynomials of the fields that are not necessarily analytic in the constant ghosts.
\section{TYM as a variable redefinition }
After twisting physical nature of some fields are interpreted differently, i.e. some fields become ghosts while some others become anti-ghosts \cite{witten}. In order to derive the topological fields that have the correct dimensions and ghost numbers via field redefinitions the aforementioned non-analyticity argument can be used.  Since the SUSY ghosts $\x_i$ have ghost number one and dimension 1/2, by studying the structure of the gauge fermion $\Psi_E$ as given in (8), the only consistent field redefinitions that assign the correct dimensionality and ghost number to the topological fields\cite{ben} are found to be
\be
A_\mu = A_\mu\quad , \quad\psi_\mu = - \x_i e_\mu \labar^i\quad , \quad\Phi=i\kok \x_i\x^i N \quad , \quad \bar{\Phi}= \frac{i}{\kok\x_i \x^i} M
\ee
\be
\eta = \frac{1}{\x _k \x^k} \x_i\la^i \quad , \quad \X_{\mu\nu}= \frac{-2}{\x _k \x^k} \x^i e_{\mu\nu} \la_i
\quad , \quad
B_{\mu\nu}= \frac{-2}{\x _k \x^k} \x^i e_{\mu\nu}\vt _i ^j \x_j . \vD
\ee

It is straightforward to show that when the above variable redefinitions are inserted in the transformations\footnote{Here,  $F_{\mu\nu}^+ = F_{\mu\nu}+\frac{1}{2}\e_{\mu\nu\la\rho}F_{\la\rho}$ is the self-dual part of the field strength $F_{\mu\nu}$.} (5), 
$$
\bs A_{\mu}= D_{\mu}c + \Psi_{\mu} \quad ,\quad\bs \psi_\mu  = - \{ c, \Psi_\mu \}  -  D_{\mu} \Phi 
\quad ,\quad\bs \Phi = - [c,\Phi ]\quad ,\quad\bs c = - \frac{1}{2}\{ c,c\} +\Phi
$$
$$
\bs \bar{\Phi} = - [c,\bar{\Phi} ] + \eta \quad ,\quad
\bs \eta = - \{c, \eta \} + [\Phi ,\bar{\Phi} ]\quad ,\quad
\bs \X_{\mu\nu} = -[c,\X_{\mu\nu} ] + F_{\mu\nu}^+ + \B_{\mu\nu}
$$
\be
\bs B_{\mu\nu} = - [c, B_{\mu\nu} ] + [\Phi , \X_{\mu\nu} ] - (D_\mu \psi_\nu -D_\nu \psi_\mu)^+
\ee
\noindent one can exactly extract the scalar supersymmetry transformations $\d$ introduced by Witten \cite{witten} if one decomposes $\bs$ on the fields $(A_{\mu},\, \Phi, \,\bar{\Phi},  \,\psi_{\mu}, \,\eta, \,\X_{\mu\nu})$ as $ \bs=s_o + \d $ \footnote{Note  that this scalar SUSY generator can also be written as a composite generator, $\d=-i\x^i Q_ i $ where $Q_i$ are the chiral SUSY generators.}.

Similarly, the corresponding action that can also be found by these field redefinitions
\bea
I_{top} &=& \bs \Psi_{top} =\bs Tr \int d^4 x  (-\yarim \eta  [\Phi ,\bar{\Phi} ] + \frac{1}{8}\X_{\mu\nu}  F_{\mu\nu}^+
-\frac{1}{8}\X_{\mu\nu}B_{\mu\nu} + \bar{\Phi} D_\mu \psi_\mu )\\
&=&Tr\int d^4 x (\frac{1}{8} F_{\mu\nu}^+ F_{\mu\nu}^+ + \eta D_\mu \psi_\mu -\frac{1}{4} \X_{\mu\nu}(D_\mu \psi_\nu -
D_\nu \psi_\mu)^+ -\bar{\Phi}D^2 \Phi-\nonumber\\
 &&\qquad - \,\,
 \! \yarim \Phi \{\eta ,\eta \} -\frac{1}{8} \Phi \{ \X_{\mu\nu}, \X_{\mu\nu}\} + \bar{\Phi} \{\psi_\mu ,
\psi_\mu  \} -\yarim [\Phi ,\bar{\Phi}]^2 -\frac{1}{8} B_{\mu\nu}B_{\mu\nu} ) .
\eea
\noindent is exactly the Topological Yang Mills action  \cite{witten} with an auxiliary field term . We remark that the inclusion of the auxiliary field is crucial in order to write the action as an exact term\footnote{The reason why the action could not be written as an exact term in the original  paper \cite{witten} is that the twisted theory was obtained from  the on-shell SYM. Note that, since $\Psi_{top}$ is gauge invariant, we have $I_{top}=\bs\Psi_{top}=\d\Psi _{top}$.}.

In other words,  TYM theory in flat Euclidean space can be obtained directly as variable redefinitions from the ordinary N=2 SYM theory \cite{ben}. As it is obvious from the above definitions of the topological fields, the ghost numbers and the dimensions that are assigned to the fields in the twisting procedure  by hand, appears here naturally due to the composite structure of the topological fields in terms of global ghosts $\x_i$ and the original fields i.e. with respect to the power of $\x_i$'s in the definitions.
\section{Baulieu-Singer formulation of TYM}
Aiming to incorporate the instantons into supersymmetric theories Zumino have constructed a supersymmetric field theory directly in the Euclidean space \cite{zumino}. It is then observed by Zumino that when one imposes for instance an anti self-dual field strength, i.e. $F_{\mu\nu}^+ =0$
with the restrictions $ M=\la_i=0$ the equations of motion from (1) reduce to a simple form \cite{zumino},
\be
F_{\mu\nu}^+ = F_{\mu\nu}+\frac{1}{2}\e_{\mu\nu\la\rho}F_{\la\rho} = 0\, ,\,D^2 N = \frac{i\kok}{2}\{ \labar^i , \labar_i \} \, ,\,
e_\mu D_\mu \labar^i = 0 .
\ee
that are invariant under the corresponding truncated SUSY.

The equations (15) are also the saddle point equations in  the context of constraint instanton method \cite{dkm}. On the other hand, similar equations are
obtained in Baulieu-Singer formulation of TYM without any approximation \cite{bftt2}. Since, both of the approaches to the instanton calculations give the same result \cite{bftt2} and Wittens TYM \cite{witten} can be obtained by using simple field redefinitions (10,11), it is natural to look for another analogy between the above instanton inspired truncation of Euclidean N=2 SYM theory and the Baulieu-Singer approach to TYM.

Indeed, when the above instanton inspired truncation is used to define another nilpotent operator $\ts$,
\be
\ts = \bs|_{F_{\mu\nu}^+ = \Ds \labar^i =M = \la_i = 0} \,\,\, ,\, \ts^2 =0
\ee
such that
\be
\ts A_{\mu}=  D_{\mu}c - \x_i e_{\mu}\labar^i \, ,\,\ts \labar^i = - \{ c,\labar^i \}  +i \kok \bar{e}_{\mu}\x^i D_{\mu} N 
\ee
\be
\ts N = - [c, N ]\, ,\,\ts c = - \frac{1}{2}\{ c,c\} + i\kok\x_i\x^i N
\ee
and\footnote{The reason why we do not set $\la_i = \vD =0$ in Eq.(20) is that the pairs $(M,\x^i\la_i)$ and $(\x^i \vt_i^j\la_j,\vD)$ behaves like the trivial pairs (BRST doublets) It is known that the cohomology of an operator does not depend on inclusion of such trivial pairs (see for instance \cite{gps,psbook}).}
\be
\ts M =  i\kok \x^i\la_i \, ,\, \ts \la_i =  \vt_i ^j \x_j.\vD \, ,\,\ts \vD = 0
\ee
$\ts$-transformations are found to be exactly that of Baulieu-Singer \cite{bs} after performing the field redefinition given in (10,11) \cite{ben}.

On the other hand the gauge fermion that is compatible with the restrictions of Zumino \cite{zumino} has to be chosen slightly different then the one given for Euclidean case (8),
\be
\Psi_{inst.} = \frac{1}{\x _k \x^k} Tr \int d^4 x (  - \frac{\a}{2} \x^i \vt _i ^j \la_j . \vD  - \frac{1}{2} \x^i
e_{\mu\nu}\la_i F_{\mu\nu}^+  + \frac{i\kok}{2}  \x^i e_\mu  \labar_i D_\mu M ) 
\ee
so that the corresponding action is
\bea
&&I_{inst.}^{(\a)}=\ts\Psi_{inst.}\\
&&=Tr \int d^4 x (-\frac{\a}{8}B_{\mu\nu}B_{\mu\nu} +\frac{1}{4}B_{\mu\nu}  F_{\mu\nu}^+  
- \la^i e_\mu D_\mu \labar_i
+ M (D_{\mu} D_{\mu}N -\frac{i\kok}{2}\{\labar^i ,\labar_i \}) +\nonumber\\
&&+  \frac{1}{\x _k \x^k} (- \frac{1}{2}\x^i e_{\mu\nu}\la_i [c,F_{\mu\nu}^+] 
+ \frac{i\kok}{2} M \{ c ,\x^i e_\mu D_\mu\labar_i \}) \,\,) 
+  \frac{1}{\x _k \x^k} Tr\int d^4 x \del_\mu( \ts \frac{i\kok}{2}M \x^i e_\mu  \labar_i )\nonumber
\eea
where we have used the definition of $B_{\mu\nu}$ in order to have notational simplification.

First of all, the gauge fermion $\Psi_{inst}$ (20) and the above action $I_{inst}$ are exactly the ones given in Baulieu-Singer approach \cite{bs} up to ordinary gauge fixing. However, if the above relations are considered on their own, to be able to derive the instanton equations (15) from the action functional without having any dependence on the constant ghosts, the coefficient
of $Tr \x^i\la_i [M,N]$ in the Euclidean $\Psi_E$ has to be chosen to vanish whereas the coefficient $\a$ of $Tr \ts \x^i \vt _i ^j \la_j . \vD$ can be left arbitrary. Therefore, the gauge fermion $\Psi_{inst}$ is the only consistent choice up to total derivatives that gives the right action to derive the exact instanton equations, when the truncated transformations (17-19) are used.\\

\noindent\textbf{Acknowledgments:} I am grateful to the organizers of the Workshop SQS'05 for inviting me and for the kind hospitality during the workshop. 

%
%
%
%

\end{document}